\documentclass[10pt,twocolumn,letterpaper]{article}

\usepackage[pagenumbers]{cvpr} 

\usepackage{booktabs}
\usepackage{amsmath}
\usepackage{amssymb}
\usepackage{multirow}

\usepackage{microtype}

\definecolor{cvprblue}{rgb}{0.21,0.49,0.74}
\usepackage[pagebackref,breaklinks,colorlinks,allcolors=cvprblue]{hyperref}

\def\confName{CVPR}
\def\confYear{2027}

\title{Secret-Stego Dissimilarity as a Design Axis: Invertible Coverless Image Steganography with Diffusion Models}

\author{
    Hongxin Xu$^{1,2}$ \quad
    Jian-Ping Mei$^{1}$ \quad
    Can Wang$^{2}$ \quad
    Defang Chen$^{3}$ \\[0.5em]
    $^{1}$Zhejiang University of Technology \quad
    $^{2}$Zhejiang University \quad
    $^{3}$UC Berkeley \\
}

\begin{document}
\maketitle

\begin{abstract}
Coverless image steganography (CIS) synthesizes a stego image rather than modifying an existing cover image, enabling authorized recipients to reconstruct the original secret image from the stego.
Existing diffusion-based CIS methods can generate natural-looking stego images but preserve substantial visual similarity to the secret image. This resemblance risks exposing structural and semantic cues, giving rise to security vulnerabilities that cannot be evaluated solely via recovery fidelity.
Achieving substantial visual dissimilarity between the secret and stego images without compromising stego quality and recovery fidelity remains challenging. To address this issue, we propose InvCISD, an invertible diffusion framework that couples the latent representations of the secret and an irrelevant reference image with an invertible network called LIMNet. We first train LIMNet in diffusion latent space, followed by end-to-end fine-tuning of the entire network, i.e., LIMNet integrated diffusion inversion and generation modules.
Experiments demonstrate that the proposed method substantially reduces secret-stego visual similarity, improves stego quality, and retains satisfactory secret reconstruction quality. Our further investigation shows that all evaluated methods are highly detectable by the CIS-oriented steganalysis model, indicating that resistance against targeted steganalysis constitutes a critical direction for future CIS research.
\end{abstract}

\section{Introduction}

Image steganography conceals secret information within an innocuous stego image, enabling its use in secure communication and privacy protection~\cite{song2024survey,raj2026comprehensive}. In image steganography, the secret may itself be a full image, so the transmitted stego image must remain visually plausible while preserving the spatial and semantic information needed for recovery. Recent deep learning and invertible-network approaches have improved the capacity~\cite{lu2021large}, the quality of reconstruction~\cite{guan2023deepmih}, and the robustness~\cite{xu2022robust} of cover-based image hiding.

\begin{figure}[t]
    \centering
    \includegraphics[width=\columnwidth]{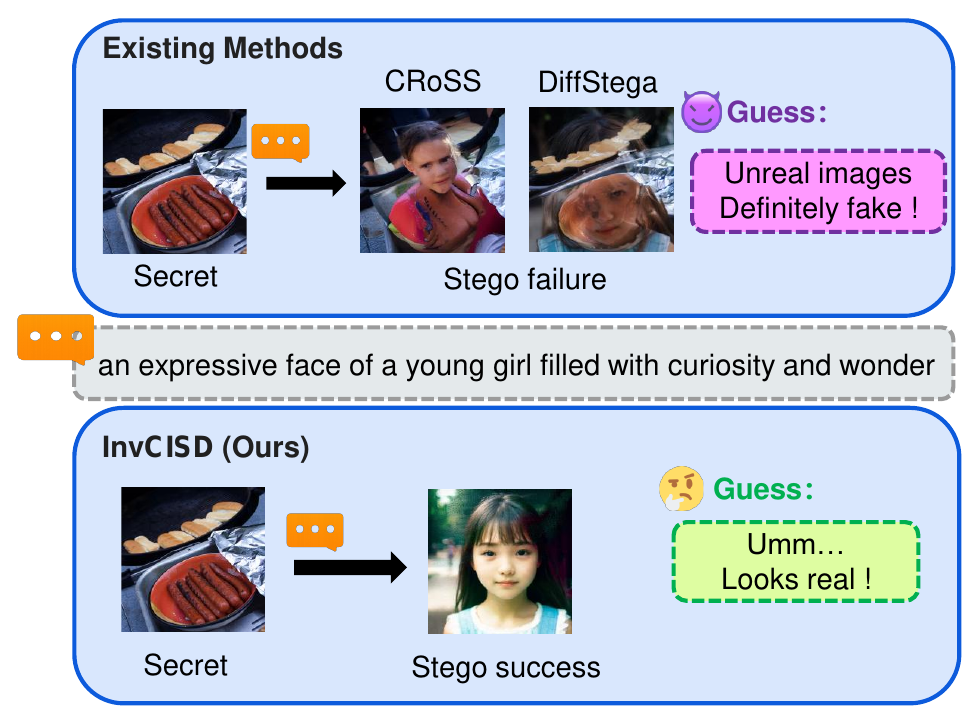}
    \caption{Visible secret leakage. \textbf{Top}: stego images generated by existing diffusion-based CRoSS~\cite{yu2023cross} and DiffStega~\cite{yang2024diffstega} retain recognizable structural and semantic secret cues. \textbf{Bottom}: the proposed InvCISD achieves high stego quality with considerable secret-stego dissimilarity.}
    \label{fig:motivation}
\end{figure}

Coverless image steganography (CIS) creates a stego image without a designated cover image~\cite{zhou2015coverless}. Compared with conventional cover-based steganography, CIS reduces the risk of hidden message detection due to cover leakage. Diffusion-based CIS leverages the strong generative capability of pretrained diffusion models~\cite{ho2020denoising,rombach2022high} to produce plausible stego images with recoverable full-image secrets~\cite{yu2023cross,zhang2026review}.
In these approaches, stego images are synthesized by editing the diffusion latent of the secret, conditioned on a reference image generated from a text prompt. To preserve reconstruction quality, secret-relevant conditioning is typically used to induce only local or subtle modifications to the secret latent during stego generation. This yields a high visual similarity between the stego and the secret, raising the risk of secret leakage upon direct visual inspection of the stego. Attempts to enlarge this visual discrepancy by adopting reference images with a greater content divergence from the secret will instead produce unnatural stego images where secret information is prominent, as illustrated in Figure~\ref{fig:motivation}. Although reduced visual similarity between secret and stego images has been identified as an important aspect of image hiding in~\cite{yang2024diffstega}, its design and evaluation schemes remain primarily centered on generation fidelity and secret recovery quality.

In this work, we explicitly incorporate the secret-stego dissimilarity into the design of diffusion-based CIS. To generate a stego with large visual difference from the secret, we propose to map the secret to an unrelated reference within the diffusion latent space. The reference can be located far from the secret's original latent representation and governs the appearance of the generated stego image. This latent mapping is required to be invertible to simultaneously ensure reliable recovery. Combining the above two considerations, we come up with an \underline{inv}ertible \underline{c}overless \underline{i}mage \underline{s}teganography framework with \underline{d}iffusion models, abbreviated as InvCISD.
We propose an effective strategy to train the invertible network (LIMNet) and provide theoretical analysis to elaborate the formulation rationale of the auxiliary latent regularizer.

Experimental results show that InvCISD achieves substantially greater secret-stego dissimilarity while maintaining satisfactory stego and secret reconstruction quality.

\noindent\textbf{Contributions:}
\begin{itemize}
    \item We formulate a challenging objective for coverless image steganography (CIS): increasing secret-stego dissimilarity while retaining high-quality stego generation and secret reconstruction.
    \item To achieve this objective, we propose LIMNet, an invertible latent mapping network, and design a training pipeline with tailored loss functions to facilitate its effective optimization and seamless integration into pre-trained diffusion backbones.
    \item We conduct comprehensive experimental evaluations of our proposed CIS approach with a dedicated experimental protocol and standardized evaluation metrics, covering five core aspects.
\end{itemize}

\section{Related Work}

\paragraph{Cover-based Image Steganography.}
In cover-based image steganography, secret messages are embedded via subtle modifications to a selected cover image. Early methods modify spatial or frequency-domain coefficients, such as least significant bits or transform coefficients. Deep learning shifted the problem toward end-to-end encoder-decoder hiding~\cite{baluja2017hiding,zhu2018hidden}. Invertible neural networks (INNs) are particularly attractive for image secrets as the embedding and extraction paths can share a bijective mapping.

Several works~\cite{jing2021hinet,lu2021large,xu2022robust,guan2023deepmih} demonstrate that invertible architectures can improve reconstruction consistency and capacity. Despite promising performance, cover-based approaches suffer from elevated detection risks caused by cover leakage~\cite{karampidis2018review,boroumand2018deep}.

\paragraph{Diffusion-based CIS.}
Diffusion models~\cite{ho2020denoising,song2021denoising} and latent diffusion models~\cite{rombach2022high,podell2024sdxl} offer strong image priors. CRoSS~\cite{yu2023cross} uses diffusion generation for controllability, robustness, and security-oriented hiding objectives, while DiffStega~\cite{yang2024diffstega} studies training-free coverless hiding with diffusion models. RoSteALS~\cite{bui2023rosteals} shows that autoencoder latent spaces can also support robust steganography. While these works motivate the use of generative latent spaces for CIS, they rarely prioritize secret-stego visual dissimilarity in their core design.

\paragraph{Diffusion Inversion and Conditioning.}
Our system relies on diffusion inversion to map observed images into latent trajectories. EDICT~\cite{wallace2023edict} improves the inversion consistency through coupled transformations. IP-Adapter~\cite{ye2023ip} provides a lightweight image prompt mechanism for text-to-image diffusion models. These off-the-shelf models and their enhanced future variants enable effective integration of invertible latent hiding with pre-trained diffusion decoders.

\section{Method}

\subsection{Overview}

\begin{figure*}[t]
    \centering
    \includegraphics[width=\textwidth]{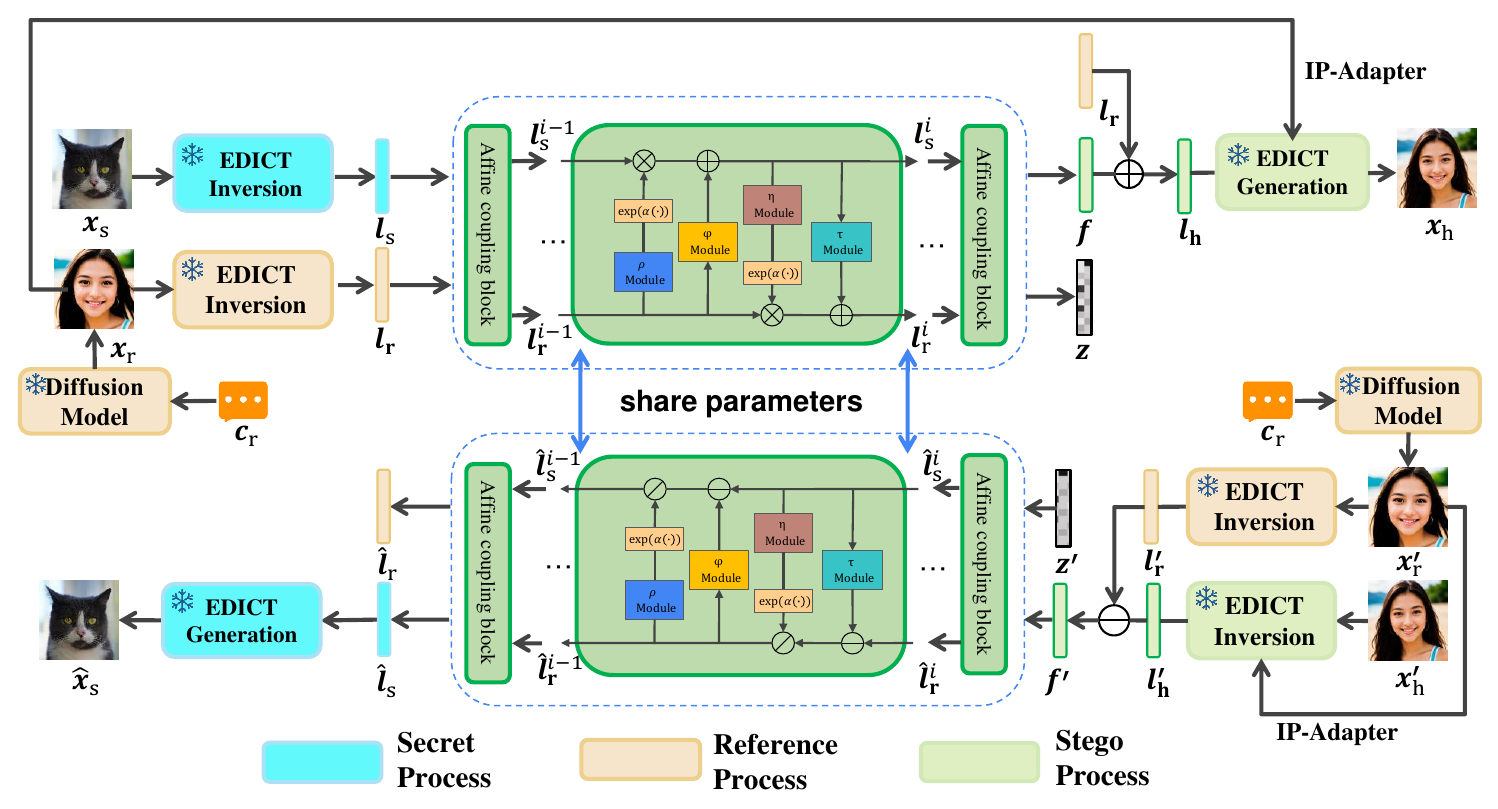}
    \caption{Overview of the proposed invertible diffusion steganography (InvCISD) framework. In the hiding stage (\textbf{top}), the secret $x_s$ and a reference $x_r$ (generated with sender-receiver shared condition $c_r$) are first encoded into diffusion latents $\ell_s$ and $\ell_r$, and then maps to a secret-bearing residual $f$ and an auxiliary latent $z$ with LIMNet; adding $f$ to $\ell_r$ yields the stego latent $\ell_h$, which gives $x_h$ through diffusion generation conditioned on $x_r$ via IP-Adapter. The recovery process is inverse to the embedding process (\textbf{bottom}): the receiver obtains $\ell_h'$ and $\ell_r'$ from the received $x'_h$ and regenerated $x'_r$, computes $f'=\ell_h'-\ell_r'$, and uses the same LIMNet with a sampled $z'$ to recover $\hat{\ell}_s$ and then $\hat{x}_s$.}
    \label{fig:method}
\end{figure*}

Given a secret image $x_s$, our goal within the diffusion-based CIS framework is to embed it into the generated stego image $x_h$ that perceptually differs from $x_s$ while enabling authorized receivers to recover an approximate secret image $\hat{x}_s$. The sender and receiver deterministically generate a reference image $x_r$ using a shared prompt and random seed. The reference provides the visible basis for $x_h$, whereas an invertible latent mapping preserves the information required for secret recovery. As illustrated in Figure~\ref{fig:method}, the framework contains a hiding stage and a recovery stage. We propose a two-stage training for learning and integrating the latent mapping network (LIMNet) with pretrained diffusion inversion and generation networks.

\paragraph{Recovery assumptions.}
Only $x_h$ is transmitted during the runtime phase. To regenerate $x_r$ and recover the secret, the receiver uses the same prompt, random seed, pretrained diffusion components, as well as inversion and generation configurations used by the sender.

\subsection{Latent Invertible Mapping}

To resolve the secret-stego dissimilarity issue under the recovery-fidelity constraint, we propose coupling the secret and a distant reference in the latent space of a pre-trained diffusion model. Specifically, we leverage EDICT~\cite{wallace2023edict} to invert $x_s$ and $x_r$ into diffusion latents $\ell_s$ and $\ell_r$, and connect them with an invertible network LIMNet, i.e., $F_\phi$.

\paragraph{Invertible architecture.}
LIMNet consists of eight affine coupling blocks~\cite{dinh2016density}. Each block uses four residual dense subnetworks for its scale and translation branches, and each residual dense subnetwork has five convolutional layers with dense connections.
Given the input and output $[\ell_s^{i-1},\ell_r^{i-1}]$ and $[\ell_s^i,\ell_r^i]$, the $i$-th block first updates the secret branch using the reference branch and then updates the reference branch using the resulting secret branch (Figure~\ref{fig:method}):
\begin{equation}
\begin{aligned}
\ell_s^i &= \ell_s^{i-1}\odot\exp\!\big(\alpha(\rho_i(\ell_r^{i-1}))\big) + \varphi_i(\ell_r^{i-1}),\\
\ell_r^i &= \ell_r^{i-1}\odot\exp\!\big(\alpha(\eta_i(\ell_s^i))\big) + \tau_i(\ell_s^i).
\end{aligned}
\label{eq:affine-forward}
\end{equation}
Here, $\rho_i$ and $\eta_i$ produce the scale signals, $\varphi_i$ and $\tau_i$ produce the translation signals, and $\alpha(\cdot)$ is the clamping function applied before exponentiation. Because each update conditions only on the other branch, the inverse is obtained in closed form and in the reverse order:
\begin{equation}
\begin{aligned}
\hat{\ell}_r^{i-1} &= \big[\hat{\ell}_r^i-\tau_i(\hat{\ell}_s^i)\big]\odot\exp\!\big(-\alpha(\eta_i(\hat{\ell}_s^i))\big),\\
\hat{\ell}_s^{i-1} &= \big[\hat{\ell}_s^i-\varphi_i(\hat{\ell}_r^{i-1})\big]\odot\exp\!\big(-\alpha(\rho_i(\hat{\ell}_r^{i-1}))\big).
\end{aligned}
\label{eq:affine-inverse}
\end{equation}
Thus, the eight-block composition maps $[\ell_s,\ell_r]$ to $[f,z]$. The same subnetworks are reused for the recovery path without requiring a separate network.

\paragraph{Hiding stage.}
Given the latent pair $[\ell_s,\ell_r]$, LIMNet produces a secret-bearing residual $f$ and an auxiliary latent $z$:
\begin{equation}
    [f, z] = F_\phi([\ell_s, \ell_r]).
\end{equation}
We construct the stego latent by adding the residual to the reference latent,
\begin{equation}
    \ell_h = \ell_r + f,
\end{equation}
and decode $\ell_h$ to produce the stego image $x_h$ with the latent diffusion model, conditioned on the IP-Adapter feature~\cite{ye2023ip} extracted from $x_r$.
This additive construction endows the two components with distinct roles: $\ell_r$ anchors the visible content of the stego image, whereas $f$ carries the information used by the inverse mapping.

The prompt and seed determine the reference content, and we use an IP-Adapter with editing strength $s$ to control the guidance strength of the reference on stego, thereby adjusting secret-stego dissimilarity. It controls the fraction of the inversion-editing trajectory used for encoding and decoding.

\paragraph{Recovery stage.}
We denote $x'_h$ as the received stego image, which may slightly differ from $x_h$ due to transmission noise. The receiver first regenerates $x'_r$ from the shared prompt and seed, and then inverts $x'_h$ and $x'_r$ to obtain $\ell_h'$ and $\ell_r'$. The received residual is computed as $f'=\ell_h'-\ell_r'$. Because transmitting $z$ would add an additional payload and risk, the receiver instead samples $z'\sim\mathcal{N}(0,I)$ and applies the inverse mapping,
\begin{equation}
    [\hat{\ell}_s,\hat{\ell}_r] = F_\phi^{-1}([f',z']).
\end{equation}
Finally, diffusion generation and VAE decoding of $\hat{\ell}_s$ yields the reconstructed secret $\hat{x}_s$. Using a sampled $z'$ means that recovery is not the exact inverse of the embedding-time mapping, even when $x'_h=x_h$. The following analysis therefore regularizes $z$ to limit the error introduced by this substitution.

\subsection{Two-Stage Training}

\paragraph{LIMNet training.}
To avoid backpropagating through the computationally expensive diffusion process, we first train LIMNet with cached latent pairs to optimize the regularized latent reconstruction loss
\begin{equation}
    \mathcal{L}_{\mathrm{LIM}} =
    \lambda_{\mathrm{rec}}\|\hat{\ell}_s-\ell_s\|_2^2
    + \lambda_z \mathcal{R}_z(z), \label{eq:stage1_loss}
\end{equation}
where
\begin{equation}
    \mathcal{R}_z(z) = \frac{1}{2d}\|z\|_2^2,
    \label{eq:auxiliary_norm}
\end{equation}
is the regularization term and $d=\dim(z)$. We show in the next section why the above formulation directly controls the expected error arising from the substitution of $z$ by $z'$.

\paragraph{End-to-end fine tuning.}
The second stage performs end-to-end fine-tuning of the entire network, which integrates LIMNet with diffusion inversion and generation.
In addition to the regularized latent reconstruction, it penalizes the difference between the decoded stego image and its reference:
\begin{equation}
\begin{aligned}
    \mathcal{L}_{\mathrm{e2e}} =
    &\lambda_{\mathrm{ref}}\|x_h-x_r\|_2^2
    + \lambda_{\mathrm{rec}}\|\hat{x}_s-x_s\|_2^2 \\
    &+ \lambda_z \mathcal{R}_z(z).
\end{aligned}
\end{equation}
This fine-tuning stage accounts for the errors introduced by diffusion inversion and image generation, which are absent from latent-only training.

\subsection{Auxiliary Latent Regularization}

We now explain why incorporating $\mathcal{R}_z(z)$ in Eq.(\ref{eq:stage1_loss}) improves recovery when the receiver samples $z'\sim\mathcal{N}(0,I)$ as an approximation of $z$. The argument applies on the latent domain encountered during training and inference. We assume that the scale and translation subnetworks in each coupling block are Lipschitz continuous, that the scale output is bounded by the clamped-exponential branch used in our implementation, and that the corresponding inputs and intermediate activations lie in a bounded domain. Under these conditions, the inverse mapping $G=F_\phi^{-1}$ is $L_G$-Lipschitz on that bounded domain for some finite $L_G$ that depends on the domain radius.

For $[f,z']$ in the same domain, if the residual is preserved, exact invertibility gives $G([f,z])=[\ell_s,\ell_r]$. The error due only to replacing $z$ with $z'$ is therefore bounded by
\begin{equation}
    \|\hat{\ell}_s-\ell_s\|
    \le \big\|[\hat{\ell}_s,\hat{\ell}_r]-[\ell_s,\ell_r]\big\|
    \le L_G\,\|z-z'\|
\end{equation}
on the event that both $[f,z]$ and $[f,z']$ lie in the applicable bounded domain.  For $z'\sim\mathcal{N}(0,I)$, this event holds with high probability: by Gaussian concentration, the sampled auxiliary latent $\|z'\|$ concentrates near $\sqrt{d}$ and lies outside a fixed radius only with exponentially small tail probability (see supplementary material).  The unconditional expectation $\mathbb{E}_{z'}\|z-z'\|_2^2 = \|z\|_2^2+d$ therefore dominates the expected squared error.  Substituting Eq.~(\ref{eq:auxiliary_norm}) gives
\begin{equation}
    \mathbb{E}_{z'}\|z-z'\|_2^2 = 2d\,\mathcal{R}_z(z)+d.
\end{equation}
Thus, $\mathcal{R}_z$ is the normalized, trainable part of the expected squared substitution error. Minimizing it directly limits the error caused by replacing $z$ with $z'$. Full statements and proofs are in the supplementary material.

\section{Experiments}

\begin{figure*}[t]
    \centering
    \includegraphics[width=\textwidth]{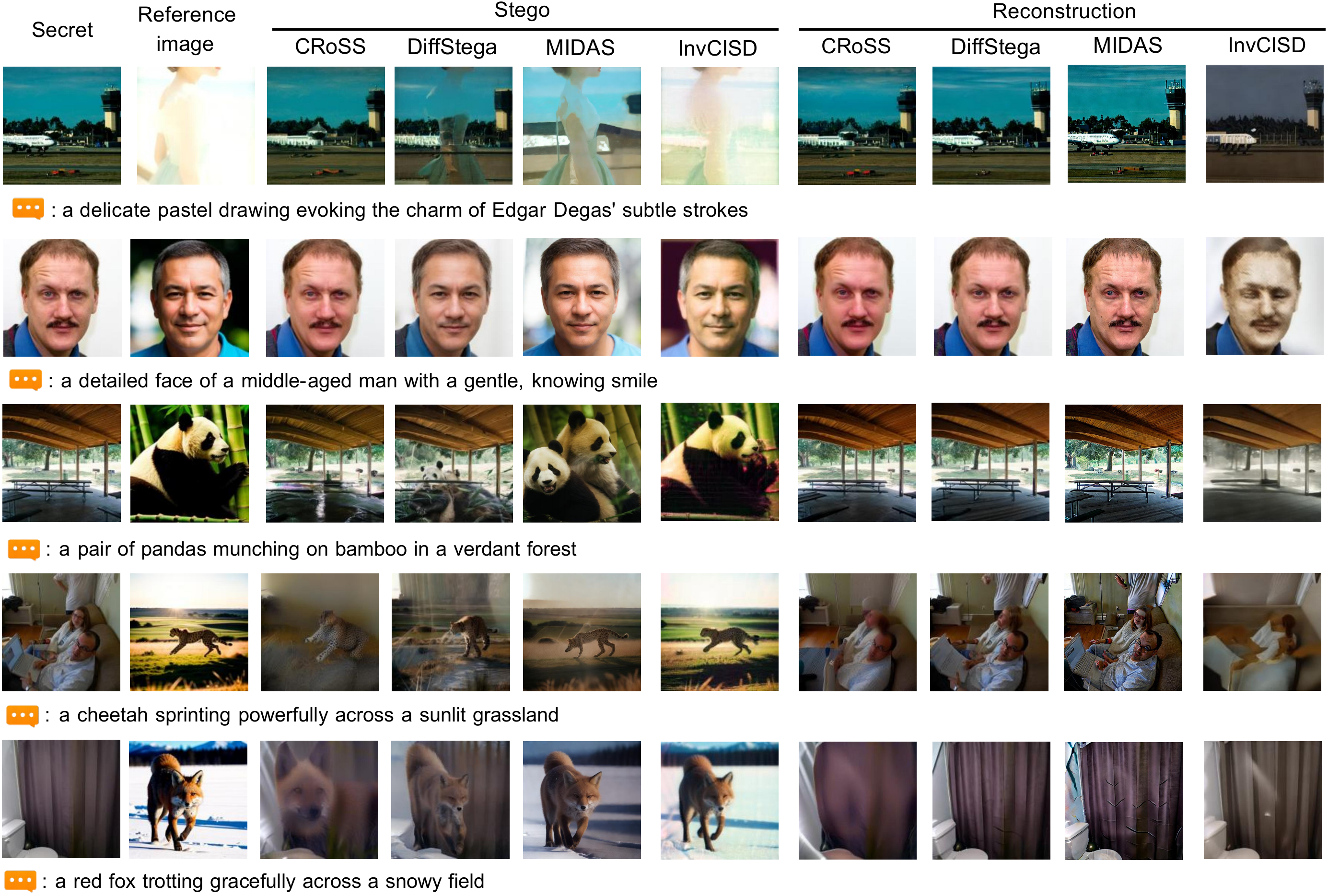}
    \caption{Visual comparison of CRoSS, DiffStega, MIDAS, and our InvCISD. Each target prompt produces a reference image semantically distant from the secret, and is used as the generation target across all four methods.}
    \label{fig:qual}
\end{figure*}

We evaluate six aspects of the proposed diffusion-based coverless image steganography approach:
\begin{itemize}
    \item visible leakage of secret information (Figure~\ref{fig:qual})
    \item secret reconstruction quality (Table~\ref{tab:main}, Figure~\ref{fig:qual}),
    \item stego quality (Table~\ref{tab:main}, Figure~\ref{fig:qual})
    \item secret-stego dissimilarity (Table~\ref{tab:main}, Figure~\ref{fig:qual}, Figure~\ref{fig:semantic_distance})
    \item steganalytic detectability (Figure~\ref{fig:detectability_combo})
    \item robustness (Table~\ref{tab:robust})
\end{itemize}

\subsection{Settings}

\paragraph{Implementation details.}
We adopt the Stable Diffusion v1.5 backbone, with IP-Adapter Plus for condition injection and EDICT as an invertible scheduler. LIMNet is trained with Adam at a learning rate of $10^{-4}$, $\beta_1=0.9$, $\beta_2=0.999$, weight decay $10^{-5}$, and a batch size of 32. Latent-space training uses $\lambda_{\mathrm{rec}}=1$ and $\lambda_z=0.5$. For end-to-end fine-tuning, we use $\lambda_{\mathrm{rec}}=1.5$, $\lambda_{\mathrm{ref}}=0.3$, and $\lambda_z=0.15$.

\paragraph{Baselines.}
We compare our proposed InvCISD against three existing approaches: CRoSS~\cite{yu2023cross}, DiffStega~\cite{yang2024diffstega}, and MIDAS~\cite{midas2026}, which follow the same diffusion-based coverless image-hiding paradigm. All four methods are evaluated using identical data splits and preprocessing pipelines. CRoSS additionally uses a per-image source caption describing the secret content as its private key. For DiffStega, we set prompt 1 to the null text and use the displayed target prompt as prompt 2. MIDAS is run in its single-image mode with the default hyperparameters of the official implementation.

\paragraph{Dataset.}
Unistega\_full contains 4,016 training pairs, 500 validation pairs, and 492 test pairs of images and prompts drawn from COCO~\cite{lin2014coco} and UniStega~\cite{yang2024diffstega}; all images are 512×512. Each pair carries a source caption and two evaluation prompts (low and high tier). For LIMNet training, we cache 5,000 secret-reference latent pairs obtained by applying EDICT inversion to image pairs from the training split, and generate 250,000 stego latents from these 5,000 originals paired with 50 references via DeepMIH~\cite{guan2023deepmih}.

\paragraph{Evaluation metrics.}
We evaluate secret-stego visual dissimilarity with PSNR, SSIM~\cite{wang2004ssim}, and LPIPS~\cite{zhang2018lpips} separately, since they capture complementary pixel-level, structural, and learned-perceptual differences. Stego PSNR, SSIM, and LPIPS compare $x_h$ with the generated reference $x_r$, while reconstruction PSNR compares $\hat{x}_s$ with $x_s$. CLIPScore~\cite{hessel2021clipscore} measures alignment between $x_h$ and the generation prompt, and NIQE~\cite{mittal2013niqe} measures the naturalness of $x_h$. Unless stated otherwise, results are produced with the default editing strength $s=0.5$.

\begin{table*}[t]
\centering
\small
\begin{tabular}{l*{10}{c}}
\toprule
\multirow{2}{*}{Method} & \multicolumn{5}{c}{Stego quality} & \multicolumn{3}{c}{\shortstack{Secret-Stego Dissimilarity}} & \multicolumn{1}{c}{\shortstack{Reconstruction}} \\
\cmidrule(lr){2-6}\cmidrule(lr){7-9}\cmidrule(lr){10-10}
& PSNR $\uparrow$ & SSIM$\uparrow$ & LPIPS$\downarrow$ & CLIPScore$\uparrow$ & NIQE$\downarrow$ & PSNR$\downarrow$ & SSIM$\downarrow$ & LPIPS$\uparrow$ & PSNR $\uparrow$ \\
\midrule
CRoSS~\cite{yu2023cross} & 8.57 & 0.207 & 0.849 & 25.37 & 4.59 & 17.11 & 0.486 & 0.481 & 18.99 \\
DiffStega~\cite{yang2024diffstega} & 7.38 & 0.177 & 0.794 & 26.69 & 3.69 & 17.68 & 0.475 & 0.487 & 21.57 \\
MIDAS~\cite{midas2026} & 11.72 & 0.362 & 0.484 & 29.26 & 6.31 & 11.52 & 0.321 & 0.758 & 19.38 \\
Ours & 17.44 & 0.417 & 0.641 & 28.08 & 4.32 & 7.99 & 0.191 & 0.871 & 19.10 \\
\bottomrule
\end{tabular}
\caption{Quantitative comparison of four diffusion-based CIS methods on the Unistega\_full dataset.}
\label{tab:main}
\end{table*}

\subsection{Results}

Table~\ref{tab:main} shows that InvCISD achieves the highest secret-stego visual dissimilarity on all three metrics. Its LPIPS is $0.871$, compared with $0.481$ for CRoSS, $0.487$ for DiffStega, and $0.758$ for MIDAS, the closest competitor. At comparable reconstruction PSNR values (19.38 dB for MIDAS and 19.10 dB for InvCISD), InvCISD achieves a substantially higher secret-stego LPIPS, reflecting an improved trade-off between dissimilarity and reconstruction PSNR. For stego quality, InvCISD obtains the best PSNR and SSIM, while MIDAS obtains the best LPIPS and CLIPScore but the worst NIQE. Figure~\ref{fig:qual} provides representative examples. Together, the results show that InvCISD increases visual dissimilarity while maintaining competitive secret reconstruction.

\subsection{Ablation Study}

Table~\ref{tab:ablation} evaluates the contribution of the main design choices across all metrics. Removing LIMNet reduces stego PSNR from $17.44$ to $13.71$ dB and secret-stego LPIPS from $0.871$ to $0.708$. Reconstruction quality improves (PSNR $19.62$ dB, LPIPS $0.365$, ID similarity $0.913$), but at the cost of substantially weaker visual dissimilarity. Removing $\mathcal{R}_z$ increases stego PSNR to $20.57$ dB and slightly improves secret-stego LPIPS to $0.883$, but reduces reconstruction PSNR to $17.83$ dB and reconstruction perceptual quality (LPIPS $0.672$, ID similarity $0.686$). This confirms that $\mathcal{R}_z$ improves reconstruction quality after auxiliary-latent replacement, trading off stego quality. Replacing EDICT with DDIM lowers reconstruction PSNR by $0.15$ dB with negligible changes in other metrics. Removing IP-Adapter reduces stego quality (LPIPS $0.667$) and CLIPScore ($27.19$) while maintaining similar reconstruction performance.

\begin{table*}[t]
\centering
\footnotesize
\setlength{\tabcolsep}{4pt}
\begin{tabular}{l *{5}{c} *{4}{c} *{4}{c}}
\toprule
\multirow{2}{*}{Variant}
  & \multicolumn{5}{c}{Stego quality}
  & \multicolumn{4}{c}{Secret-Stego Dissimilarity}
  & \multicolumn{4}{c}{Reconstruction} \\
\cmidrule(lr){2-6}\cmidrule(lr){7-10}\cmidrule(lr){11-14}
& PSNR$\uparrow$ & SSIM$\uparrow$ & LPIPS$\downarrow$ & CLIP$\uparrow$ & NIQE$\downarrow$
& PSNR$\downarrow$ & SSIM$\downarrow$ & LPIPS$\uparrow$ & ID dist$\uparrow$
& PSNR$\uparrow$ & SSIM$\uparrow$ & LPIPS$\downarrow$ & ID sim$\uparrow$ \\
\midrule
default
  & 17.44 & 0.417 & 0.641 & 28.08 & 4.32
  & 7.99  & 0.191 & 0.871 & 0.314
  & 19.10 & 0.516 & 0.624 & 0.768 \\
w/o LIMNet
  & 13.71 & 0.320 & 0.678 & 27.21 & 4.04
  & 9.68  & 0.292 & 0.708 & 0.226
  & 19.62 & 0.562 & 0.365 & 0.913 \\
w/o $\mathcal{R}_z$
  & 20.57 & 0.574 & 0.616 & 29.89 & 3.98
  & 7.68  & 0.182 & 0.883 & 0.332
  & 17.83 & 0.445 & 0.672 & 0.686 \\
DDIM$^{\dagger}$
  & 17.46 & 0.421 & 0.638 & 28.01 & 4.39
  & 7.98  & 0.191 & 0.873 & 0.310
  & 18.95 & 0.515 & 0.626 & 0.778 \\
w/o IP-Adapter$^{\dagger}$
  & 17.31 & 0.410 & 0.667 & 27.19 & 4.27
  & 7.97  & 0.191 & 0.875 & 0.321
  & 19.06 & 0.512 & 0.632 & 0.748 \\
\bottomrule
\end{tabular}
\caption{Ablation results of InvCISD. Columns marked $\dagger$ are inference-time ablations on the same checkpoint.}
\label{tab:ablation}
\end{table*}

\begin{figure}[t]
    \centering
    \includegraphics[width=\columnwidth]{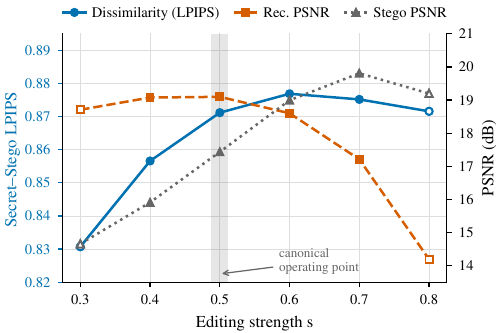}
    \caption{Effect of editing strength on secret-stego visual dissimilarity and reconstruction quality.}
    \label{fig:tradeoff}
\end{figure}

\subsection{Effect of Editing Strength}

Editing strength controls the trade-off between secret-stego dissimilarity and secret reconstruction quality. Figure~\ref{fig:tradeoff} visualizes the trend and Table~\ref{tab:tradeoff} reports full metrics across editing strengths $s \in \{0.3, 0.4, 0.5, 0.6, 0.7, 0.8\}$ on a 100-image subset. Increasing $s$ from $0.4$ to $0.5$ raises the secret-stego LPIPS from $0.857$ to $0.871$ while reconstruction PSNR remains nearly unchanged ($19.08$ to $19.10$ dB). Stego PSNR improves from $15.91$ to $17.44$ dB. Increasing $s$ to $0.6$ provides a smaller LPIPS gain of $0.006$ but lowers reconstruction PSNR by $0.50$ dB and degrades reconstruction perceptual quality (LPIPS increases from $0.624$ to $0.680$, ID similarity drops from $0.768$ to $0.721$). Beyond $s=0.6$, reconstruction quality deteriorates further while stego quality plateaus and dissimilarity gains diminish. We therefore use $s=0.5$ as the default setting, balancing dissimilarity, reconstruction, and stego quality.

\begin{table*}[t]
\centering
\footnotesize
\setlength{\tabcolsep}{4pt}
\begin{tabular}{c *{5}{c} *{4}{c} *{4}{c}}
\toprule
\multirow{2}{*}{$s$}
  & \multicolumn{5}{c}{Stego quality}
  & \multicolumn{4}{c}{Secret-Stego Dissimilarity}
  & \multicolumn{4}{c}{Reconstruction} \\
\cmidrule(lr){2-6}\cmidrule(lr){7-10}\cmidrule(lr){11-14}
& PSNR$\uparrow$ & SSIM$\uparrow$ & LPIPS$\downarrow$ & CLIP$\uparrow$ & NIQE$\downarrow$
& PSNR$\downarrow$ & SSIM$\downarrow$ & LPIPS$\uparrow$ & ID dist$\uparrow$
& PSNR$\uparrow$ & SSIM$\uparrow$ & LPIPS$\downarrow$ & ID sim$\uparrow$ \\
\midrule
0.3 & 14.66 & 0.343 & 0.742 & 26.11 & 4.27 & 8.83 & 0.214 & 0.831 & 0.296 & 18.71 & 0.483 & 0.569 & 0.787 \\
0.4 & 15.91 & 0.375 & 0.701 & 27.25 & 4.37 & 8.44 & 0.200 & 0.857 & 0.307 & 19.08 & 0.514 & 0.583 & 0.788 \\
0.5 & 17.44 & 0.417 & 0.641 & 28.08 & 4.32 & 7.99 & 0.191 & 0.871 & 0.313 & 19.10 & 0.516 & 0.624 & 0.768 \\
0.6 & 18.99 & 0.473 & 0.570 & 28.44 & 4.16 & 7.50 & 0.183 & 0.877 & 0.317 & 18.60 & 0.503 & 0.680 & 0.721 \\
0.7 & 19.81 & 0.541 & 0.512 & 28.44 & 4.28 & 7.07 & 0.174 & 0.875 & 0.331 & 17.21 & 0.470 & 0.752 & 0.635 \\
0.8 & 19.20 & 0.574 & 0.500 & 27.77 & 5.71 & 6.70 & 0.164 & 0.872 & 0.355 & 14.19 & 0.402 & 0.850 & 0.542 \\
\bottomrule
\end{tabular}
\caption{Full metric results of InvCISD across editing strengths.}
\label{tab:tradeoff}
\end{table*}

\subsection{Steganalysis}

To evaluate detectability under controlled conditions, we construct two prompt tiers that differ in the semantic distance between the secret and the generated stego. The \emph{low tier} uses near-category substitution prompts (e.g., one dog breed replaced by another), while the \emph{high tier} uses cross-category or cross-style prompts (e.g., an animal replaced by a building).

\begin{figure*}[t]
  \centering
  \includegraphics[width=0.9\textwidth]{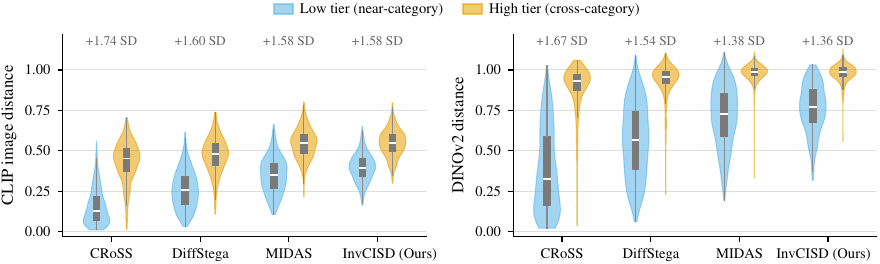}
  \caption{Secret-stego image distances for the low tier (near-category, blue) and high tier (cross-category, orange), measured with CLIP image embeddings (left) and DINOv2 features (right). Each violin summarizes the distribution over 492 paired samples for one method and tier. Both feature spaces consistently distinguish the two tiers across all evaluated methods.}
  \label{fig:semantic_distance}
\end{figure*}

\paragraph{Semantic distance verification.}
Figure~\ref{fig:semantic_distance} confirms that the tier manipulation is realized: across all four methods and two independent embedding models (CLIP and DINOv2), the high-tier distributions are shifted upward by $+1.36$ to $+1.74$ standard deviations relative to the paired low-tier values. Notably, InvCISD attains consistently larger secret-stego distances under both tiers.

\begin{figure*}[t]
  \centering
  \includegraphics[width=0.83\textwidth]{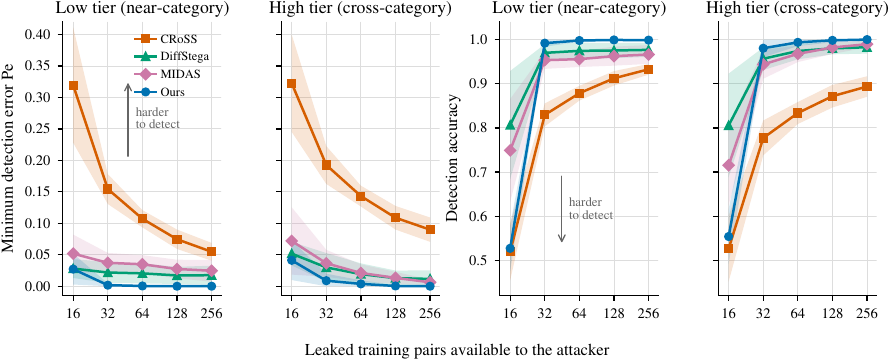}
  \caption{XuNet detection accuracy (bottom) and minimum detection error $P_e$ (top) as functions of leaked reference-stego pairs, for the low tier (left) and high tier (right). Shaded bands show $\pm$1 standard deviation across folds and seeds. All methods become detectable with sufficient attacker data.}
\label{fig:detectability_combo}
\end{figure*}

\paragraph{Detectability.}
We evaluate detectability by retraining XuNet~\cite{xu2016structural}, which is conventionally trained on cover-stego pairs. Because CIS has no pre-existing cover image, we use reference images as the clean class. We train XuNet using five-fold cross-validation, vary the number of leaked pairs available to the attacker from 16 to 256, and report results for both tiers.
Figure~\ref{fig:detectability_combo} shows detection accuracy and minimum detection error $P_e$ as functions of leaked reference-stego pairs for both the low and high tiers. All four methods become highly detectable with sufficient training data. At 256 pairs, detection accuracy exceeds $95\%$ and $P_e$ values converge across tiers for all methods.

\subsection{Robustness}

Table~\ref{tab:robust} evaluates secret reconstruction quality after common distortions. All methods are tested at their default settings. For JPEG compression at quality 90, InvCISD achieves $17.34$ dB, comparable to MIDAS ($17.26$ dB) and lower than CRoSS ($18.75$ dB) and DiffStega ($21.44$ dB) under ideal conditions. Under Gaussian noise ($\sigma=0.05$), InvCISD and MIDAS drop to $9.53$ and $10.33$ dB respectively, while CRoSS and DiffStega retain higher robustness at $17.14$ and $19.00$ dB. For rescaling to $0.5\times$, InvCISD achieves $14.37$ dB. Under $80\%$ center cropping, all methods suffer substantial degradation, with InvCISD at $12.30$ dB and MIDAS notably lower at $8.04$ dB. These results indicate that the increased visual dissimilarity achieved by InvCISD and MIDAS comes at the cost of reduced robustness to channel perturbations.

\begin{table}[t]
\centering
\small
\setlength{\tabcolsep}{5pt}

\begin{tabular}{llccc}
\toprule
Distortion & InvCISD & CRoSS & DiffStega & MIDAS \\
\midrule
JPEG ($Q=90$)     & 17.34 & 18.75 & 21.44 & 17.26 \\
Noise ($\sigma=0.05$)  & 9.53 & 17.14 & 19.00 & 10.33 \\
Resize ($0.5\times$)   & 14.37 & 18.27 & 21.42 & 18.16 \\
Crop ($80\%$)   & 12.30 & 12.40 & 12.32 & 8.04 \\
\bottomrule
\end{tabular}
\caption{Secret reconstruction PSNR (dB) after common distortions}
\label{tab:robust}
\end{table}
\section{Discussion and Limitations}

Latent regularization also shapes the reconstruction-stego trade-off. The auxiliary latent $z$ is regularized toward a small norm during training; a looser constraint gives the network more freedom to minimize stego distortion, improving stego quality at the cost of recovery quality, while a tighter constraint has the opposite effect. Empirically, weaker $z$ regularization achieved higher stego quality but lower reconstruction quality, whereas stronger regularization maintained better recovery with a moderate stego-quality penalty.

The auxiliary latent is sampled from $\mathcal{N}(0,I)$ at inference time to maintain consistency with the training regime. During data generation, the forward coupling network processes latent pairs where the auxiliary component is drawn from this distribution. The learned inverse mapping therefore expects auxiliary inputs with norm around $\sqrt{d}$, where $d$ is the latent dimension and $\mathbb{E}\|z'\|^2 = d$ for a standard Gaussian. While the $\mathcal{R}_z$ regularization encourages the network-produced latents $z$ to have small norms, the network has been trained to invert inputs following the Gaussian sampling distribution rather than other alternatives. Sampling from a distribution also preserves variability during both training and inference, supporting diverse latent-space trajectories and maintaining the expressiveness of the stego generation process. The theoretical analysis in the supplementary material formalizes how this sampling strategy bounds the expected reconstruction error under the learned distribution.

Secret reconstruction is constrained by multiple factors. First, recovery is sensitive to image perturbations because the received image must be accurately inverted into the diffusion latent space. It also requires the receiver to reproduce the prompt, seed, model checkpoints, inversion settings, and editing strength used by the sender. Finally, reconstruction is limited by both VAE encoding and multi-step inversion.

\paragraph{Limitations.}
Large editing strengths can produce stego images that are visually distant from the secret but difficult to invert, leading to blurred or semantically incomplete reconstructions. Weak reference conditioning can also increase visual separation while reducing naturalness or prompt consistency. Recovery may also fail when the sender and receiver use different prompts, seeds, checkpoints, or inversion settings, even if the received image appears natural.

The methods are evaluated at their default settings. This evaluation therefore does not identify a causal effect of visual dissimilarity on detectability. It establishes only that the evaluated methods are readily distinguishable from their generated references with the tested XuNet setting.

\section{Conclusion}

We proposed InvCISD, an invertible diffusion-based framework for coverless image steganography that explicitly addresses secret-stego visual dissimilarity. By introducing LIMNet, an invertible latent mapping network, we enable the generation of stego images that are visually distant from the secret while maintaining satisfactory reconstruction quality. Our two-stage training strategy effectively integrates LIMNet with pretrained diffusion models, and we provide theoretical justification for the auxiliary latent regularization term.

Experimental results demonstrate that InvCISD achieves substantially higher secret-stego dissimilarity compared to existing methods while maintaining competitive stego quality and reconstruction fidelity. We also conduct comprehensive steganalysis evaluations under controlled semantic-distance conditions, revealing that all evaluated methods remain highly detectable with sufficient attacker resources. This finding underscores the importance of developing resistance against targeted steganalysis as a critical direction for future CIS research.

Future work may explore adaptive editing strength selection, alternative invertible architectures, and techniques to improve robustness against steganalytic attacks while maintaining the achieved visual dissimilarity.

\section*{LLM Usage Statement}
This work was completed with the assistance of large language models for the following purposes: literature review, code debugging, writing assistance, and experimental design consultation. All core ideas, methodology design, experimental implementation, and result analysis were completed by the authors.

{\small
\bibliographystyle{ieeenat_fullname}
\bibliography{references}
}

\clearpage
\setcounter{page}{1}
\maketitlesupplementary

\appendix

\section{Reconstruction Error Bound}

This section gives the formal statements and proofs behind the main-text claim that regularizing the auxiliary latent $z$ toward $\mathcal{N}(0,I)$ bounds the secret-latent reconstruction error. Throughout, $\|\cdot\|$ is the Euclidean norm and $F_\phi=b_N\circ\cdots\circ b_1$ is the forward LIMNet built from $N$ affine coupling blocks $b_i$, with exact inverse $G=F_\phi^{-1}=g_1\circ\cdots\circ g_N$ where $g_i=b_i^{-1}$.

\textbf{Definition (Lipschitz continuity).} A map $F:\mathbb{R}^n\to\mathbb{R}^m$ is $L$-Lipschitz if $\|F(u)-F(v)\|\le L\|u-v\|$ for all $u,v$; the smallest such $L$ is its Lipschitz constant.

We repeat the three design conditions from the main text.

\textbf{Assumption 1 (Lipschitz subnetworks).} Every scale and translation subnetwork in each coupling block is Lipschitz continuous. Spectral-normalized linear layers satisfy $\|W\|_2\le1$ and hence are $1$-Lipschitz, and standard activations (ReLU, LeakyReLU, Tanh) are $1$-Lipschitz, so any finite composition of them is Lipschitz.

\textbf{Assumption 2 (bounded scale).} Each scale subnetwork output $\rho$ obeys $\|\rho(\cdot)\|_\infty\le C$ for a fixed $C>0$, so the per-channel multiplicative factor $\exp(\rho)$ lies in $[e^{-C},e^{C}]$. A clamped-exponential (or scaled-$\tanh$) activation on the scale branch enforces this.

\textbf{Assumption 3 (bounded domain).} There exist nested radii $R_N\ge R_{N-1}\ge\cdots\ge R_0>0$ such that for any $u\in\mathcal{B}_{R_i}:=\{v\in\mathbb{R}^{2d}:\|v\|\le R_i\}$, the block inverse $g_i$ maps $u$ into $\mathcal{B}_{R_{i-1}}$. Equivalently, the composition $g_1\circ\cdots\circ g_N$ maps $\mathcal{B}_{R_N}$ into $\mathcal{B}_{R_0}$ through these intermediate bounds. This is the bounded-domain condition of the main text. It is needed because the multiplicative factor $\exp(-\rho(\cdot))$ in the block inverse is input dependent: Assumption~2 bounds its magnitude, but the Lipschitz constant of the product with the unbounded translation term grows with the norm of the input.

\textbf{Lemma 1 (composition).} If $F_1$ is $L_1$-Lipschitz on a set $S$ and $F_2$ is $L_2$-Lipschitz on $F_1(S)$, then $F_2\circ F_1$ is $L_1L_2$-Lipschitz on $S$.

\textit{Proof.} $\|F_2(F_1(u))-F_2(F_1(v))\|\le L_2\|F_1(u)-F_1(v)\|\le L_1L_2\|u-v\|$. \hfill$\square$

\textbf{Lemma 2 (product with a bounded factor).} Let $u:S\to\mathbb{R}^{m}$ be $L_u$-Lipschitz with $\sup_{x\in S}\|u(x)\|\le M_u$, and let $v:S\to\mathbb{R}^{m}$ be $L_v$-Lipschitz with $\sup_{x\in S}\|v(x)\|_\infty\le M_v$. Then $x\mapsto u(x)\odot v(x)$ is $(M_uL_v+M_vL_u)$-Lipschitz on $S$.

\textit{Proof.} For $x,y\in S$, $\|u(x)\odot v(x)-u(y)\odot v(y)\|\le\|u(x)\odot(v(x)-v(y))\|+\|(u(x)-u(y))\odot v(y)\|\le\|u(x)\|\,\|v(x)-v(y)\|_\infty+\|v(y)\|_\infty\,\|u(x)-u(y)\|\le(M_uL_v+M_vL_u)\|x-y\|$, using $\|v(x)-v(y)\|_\infty\le\|v(x)-v(y)\|$. \hfill$\square$

\textbf{Lemma 3 (output bound propagation).} Let $h:\mathcal{B}_{R}\to\mathbb{R}^{2d}$ be $L$-Lipschitz with output bound $\sup_{u\in\mathcal{B}_{R}}\|h(u)\|\le M(R)$. Then $h$ maps $\mathcal{B}_{R}$ into $\mathcal{B}_{M(R)}$.

\textit{Proof.} For any $u\in\mathcal{B}_{R}$, $\|h(u)\|\le M(R)$ by hypothesis, so $h(u)\in\mathcal{B}_{M(R)}$. \hfill$\square$

\textbf{Lemma 4 (block inverse).} Under Assumptions 1--3, each block inverse $g_i:\mathcal{B}_{R_i}\to\mathcal{B}_{R_{i-1}}$ is Lipschitz continuous on $\mathcal{B}_{R_i}$ with a finite constant $L_i(R_i)$ that grows at most quadratically in $R_i$. Moreover, the output satisfies $\sup_{u\in\mathcal{B}_{R_i}}\|g_i(u)\|\le M_i(R_i)$ for some $M_i(R_i)$ affine in $R_i$, which justifies taking $R_{i-1}=M_i(R_i)$ in Assumption~3.

\textit{Proof.} Write the block input as $(x_1,x_2)=[\hat{\ell}_s^{i},\hat{\ell}_r^{i}]$ and its output as $(y_1,y_2)=[\hat{\ell}_s^{i-1},\hat{\ell}_r^{i-1}]$. By Eq.~(4) in the main paper, the block inverse first computes $y_2$ from $(x_2,x_1)$ and then $y_1$ from $(x_1,y_2)$, each by a step of the form $a=(b-t(c))\odot\exp(-\rho(c))$ with a translation subnetwork $t$ and a clamped scale subnetwork $\rho$. Apply Lemma~2 to one step with $u(b,c)=b-t(c)$ and $v(c)=\exp(-\rho(c))$ on a set where $\|b\|,\|c\|\le R_i$. There, $u$ is $(1+L_t)$-Lipschitz by Assumption~1 and $\|u\|\le R_i+\|t(0)\|+L_tR_i=:M_u(R_i)$; $v$ has entries in $[e^{-C},e^{C}]$ by Assumption~2, so $\|v\|_\infty\le e^{C}$, and $v$ is $e^{C}L_\rho$-Lipschitz because $\exp(-\,\cdot\,)$ is $e^{C}$-Lipschitz on $[-C,C]$. Hence the step is Lipschitz with constant $M_u(R_i)e^{C}L_\rho+e^{C}(1+L_t)$, affine in $R_i$, and its output satisfies $\|a\|\le e^{C}M_u(R_i)$. For the first step the input domain is $\mathcal{B}_{R_i}$ by Assumption~3; for the second step the input bound is affine in $R_i$ by the output bound of the first. Writing $g_i$ as the composition of $(x_1,x_2)\mapsto(x_1,y_2)$ and $(x_1,y_2)\mapsto(y_1,y_2)$, where the pass-through coordinate adds at most one to each Lipschitz constant, $g_i$ is Lipschitz on $\mathcal{B}_{R_i}$ by Lemma~1 with constant $L_i(R_i)$ at most quadratic in $R_i$. The final output bound $M_i(R_i)=e^{C}M_u^{(2)}(e^{C}M_u^{(1)}(R_i))$ is affine in $R_i$ because each $M_u$ is affine, so Assumption~3's requirement that $g_i:\mathcal{B}_{R_i}\to\mathcal{B}_{R_{i-1}}$ is satisfied by taking $R_{i-1}=M_i(R_i)$. \hfill$\square$

\textbf{Theorem 1 (network inverse).} Under Assumptions 1--3, the full inverse map $G=g_1\circ\cdots\circ g_N$ maps $\mathcal{B}_{R_N}$ into $\mathcal{B}_{R_0}$ and is Lipschitz continuous on $\mathcal{B}_{R_N}$ with
\begin{equation}
    L_G(R_N)\le\prod_{i=1}^{N}L_i(R_i).
\end{equation}

\textit{Proof.} By Assumption~3, each $g_i:\mathcal{B}_{R_i}\to\mathcal{B}_{R_{i-1}}$ is $L_i(R_i)$-Lipschitz by Lemma~4. The composition $G=g_1\circ\cdots\circ g_N$ maps $\mathcal{B}_{R_N}\to\mathcal{B}_{R_{N-1}}\to\cdots\to\mathcal{B}_{R_0}$, and applying Lemma~1 across the $N$-fold composition gives an overall Lipschitz constant on $\mathcal{B}_{R_N}$ no larger than $\prod_{i=1}^{N}L_i(R_i)$. \hfill$\square$

\textbf{Corollary 1 (reconstruction error).} Let $(f,z)=F_\phi([\ell_s,\ell_r])$ at embedding time, and define the event $\mathcal{E}_{R_N}:=\{[f,z]\in\mathcal{B}_{R_N}\}\cap\{[f,z']\in\mathcal{B}_{R_N}\}$ where $z'\sim\mathcal{N}(0,I)$. If the channel is ideal and $f$ is preserved, and the receiver forms $[\hat{\ell}_s,\hat{\ell}_r]=G([f,z'])$, then exact invertibility gives $G([f,z])=[\ell_s,\ell_r]$ and the expected squared reconstruction error decomposes as
\begin{equation}
    \mathbb{E}\|\hat{\ell}_s-\ell_s\|^2
    = \mathbb{E}\big[\|\hat{\ell}_s-\ell_s\|^2\mid\mathcal{E}_{R_N}\big]\,\mathbb{P}(\mathcal{E}_{R_N})
    + \mathbb{E}\big[\|\hat{\ell}_s-\ell_s\|^2\mid\mathcal{E}_{R_N}^c\big]\,\mathbb{P}(\mathcal{E}_{R_N}^c).
\end{equation}
On the event $\mathcal{E}_{R_N}$, Theorem~1 gives
\begin{equation}
    \|\hat{\ell}_s-\ell_s\|\le L_G(R_N)\,\|z-z'\|,
\end{equation}
so the conditional expected squared error is bounded by $L_G(R_N)^2\,\mathbb{E}[\|z-z'\|^2\mid\mathcal{E}_{R_N}]$. On $\mathcal{E}_{R_N}^c$, the error is bounded by the worst-case reconstruction error across the latent space (typically of order unity in the normalized latent domain).

\textit{Proof.} The decomposition is by the law of total expectation. On $\mathcal{E}_{R_N}$, both $[f,z]$ and $[f,z']$ lie in $\mathcal{B}_{R_N}$, so $\|G([f,z'])-G([f,z])\|\le L_G(R_N)\|z-z'\|$ by Theorem~1, and $\|\hat{\ell}_s-\ell_s\|\le\|[\hat{\ell}_s,\hat{\ell}_r]-[\ell_s,\ell_r]\|$ by the triangle inequality. On $\mathcal{E}_{R_N}^c$, at least one argument lies outside the applicable domain, so we use only the trivial bound. \hfill$\square$

\textbf{Remark.} The bound is an existence result under Assumptions 1--3: our implementation satisfies Assumption 2 through the clamped-exponential scale branch but does not currently enforce spectral normalization, so the result justifies the loss design—minimizing $\mathcal{R}_z$ shrinks $\mathbb{E}\|z-z'\|^2$ and hence the expected squared error bound—rather than certifying a numerical $L_G(R_N)$ for the trained network. The bound also isolates the coupling substitution and excludes diffusion-inversion and VAE-decoding error, which separately limit end-to-end fidelity.

In the implementation the clamp is $\alpha(s)=c\,(2\sigma(s)-1)$ with $c=2\times10^{-3}$, so Assumption~2 holds with $C=c$ and $e^{C}\approx1.002$. Assumption~3's nested radii can be verified empirically: for the trained network, the sequence $\{M_i(R_i)\}$ from Lemma~4 gives explicit bounds on intermediate outputs given an initial radius $R_N$.

The event $\mathcal{E}_{R_N}$ in Corollary~1 holds with high probability. For the forward component, $\|f\|$ is bounded by the range of the encoder output on training data. For the sampled $z'$, note that $\|z'\|$ is a $1$-Lipschitz function of a standard Gaussian vector, so by Gaussian concentration~\cite{vershynin2018high}, $\|z'\|\le\sqrt{d}+t$ except with probability at most $e^{-t^{2}/2}$. Similarly, $\|z\|$ concentrates near $\sqrt{d}$ when the regularization successfully drives $z$ toward $\mathcal{N}(0,I)$. Taking $R_N\ge\sup\|f\|+\sqrt{d}+t$ with $t=O(\sqrt{d})$, the tail probability $\mathbb{P}(\mathcal{E}_{R_N}^c)$ is exponentially small in $d$. The conditional expectation $\mathbb{E}[\|z-z'\|^2\mid\mathcal{E}_{R_N}]$ differs from the unconditional expectation $\|z\|^2+d$ only by terms that vanish as the tail probability vanishes, so the main-text statement $\mathbb{E}\|z-z'\|^2=\|z\|^2+d=2d\,\mathcal{R}_z(z)+d$ remains the dominant term in practice.

\end{document}